# Reacting Hydrogen Jet in Crossflow - Flame Dynamics Under Acoustic Forcing

Ankur Kumar and Anubhav Sinha*

Department of Mechanical Engineering

Indian Institute of Technology (Banaras Hindu University), Varanasi, India

*Corresponding author email – er.anubhav@gmail.com

*Abstract*

*This paper presents experimental study of reacting hydrogen jet in crossflow. High speed shadowgraph images are used to capture flame dynamics. Unforced jets with various momentum flux ratios (q) are studied. Proper Orthogonal Decomposition (POD) analysis is used to examine the high-speed instantaneous images and characterize spatio-temporal behavior the reacting jets. Further, jet is acoustically forced with various frequencies and the effect of forcing frequency and momentum flux ratio is studied in detail. Instantaneous images, POD mode shapes and Power Spectral Density (PSD) plots are used to assess the response of jet to forcing function, and the influence of crossflow. It is observed that the jet does not equally respond to all frequencies and the preferred frequency range is identified. It is also observed that higher momentum ratio produces stronger interaction with the crossflow which dampens the effect of forcing.*

*Keywords* – Hydrogen combustion, jet in crossflow, acoustic forcing, POD, flame oscillations.

## Introduction

Hydrogen is a promising alternative to fossil-fuels and the push towards clean energy and decarbonization has further accelerated the growth of hydrogen infrastructure and research for hydrogen combustion. As hydrogen has very different physical properties and combustion characteristics than general hydrocarbon fuels, the current infrastructure designed for hydrocarbon fuels needs to be adapted for hydrogen. The properties of hydrogen combustion—such as low ignition energy, high diffusivity, wide flammability limits, and fast flame speeds—pose unique challenges to controlling combustion behavior. Hydrogen is also much lighter as compared to hydrocarbon fuels. Hence, mixing and penetration of hydrogen jets will not be same as other fuels. A common fuel injection strategy is Jet in crossflow (JIC). This configuration can be found in various engineering as well as natural processes, particularly relevant to aerospace applications. In JIC, the interaction of jet with crossflow produces a range of vortex structures such as counter-rotating vortex pair (CVP), kidney vortices, hairpin vortices near the jet injection wall, etc. [1-3]. The vortex structure most relevant to the present study is that formed in the shear layer between the jet and the crossflow. The present study focuses on reacting hydrogen jet in crossflow, which has not studied extensively, especially considering the significance of the configuration in aviation and power generation applications.

The combustion characteristics of the nitrogen-diluted hydrogen jet in crossflow are numerically investigated by Murugvel et al. [4] using a LES framework. They have investigated the effect of methane addition in hydrogen jet. They explore the non-premixed nature of flame and comment on the jet mixing characteristics. The study by Sharma et al., [5] explores flame stabilization in hydrogen-enriched jet flames within a crossflow, highlighting the effectiveness of a strain-sensitive premixed model in capturing complex combustion dynamics. Hydrogen flames in in a swirl stabilized configuration is studied by Vaysse et al. [6]. They inject hydrogen in a crossflow of the swirl injector. They investigated flame stability and flame anchoring experimentally. Rasheed and Mishra [7] have studied a sonic hydrogen jet injected into a supersonic crossflow numerically. The study explores various turbulence models and compares their effectiveness in the given configuration by comparing against experimental data. Hydrogen jet with methane addition in crossflow configuration is studied by Boxx et al. [8]. They have used elevated pressure and temperature to understand combustor dynamics for gas turbine application. Local heat release form flame is understood to be aiding in flame anchoring. The study provides insights into how varying hydrogen content influences flame behavior and stabilization mechanisms, which can inform future designs of combustion systems. The same experimental facility is used for another study at realistic conditions by Saini et al. [9]. Hydrogen-enriched flames in crossflow are observed to exhibit enhanced flame-holding characteristics. The jet to crossflow momentum ratio is observed to govern the penetration and jet trajectory. They also highlight the importance of hydrogen enrichment in enhancing flame stability and influencing flow dynamics in turbulent jet flames.

A DNS study by Xu et al. [10] revealed two flame branches in the hydrogen jet: a thin windward branch and a thick leeward branch, consistent with experimental observations of hydrogen flames in crossflow. They have used OH as a marker for flame front and have also got good match with the experimental data from literature. Hydrogen flames in supersonic crossflow exhibit distinct ignition mechanisms [11], with autoignition and flame stabilization influenced by shock waves and momentum flux ratios, particularly at higher ratios. The research provides insights into the complex interactions between turbulent mixing and combustion in supersonic flows, highlighting the significant role of momentum flux ratios in determining ignition and combustion characteristics. Zhao et al. [12] have numerically investigated the hydrogen and methane jet in supersonic crossflow. The jet flames exhibit stabilization on the windward side at higher enrichment levels, with increased sootiness and greater jet penetration due to reduced crossflow entrainmen. The findings highlight the significant impact of momentum flux ratios on ignition and combustion characteristics in supersonic flows. Transverse fuel injection with hydrogen is explored experimentally by Olivani and Cozzi [13]. They observed enhanced fuel-air mixing, and reduced soot formation compared to axial injection, which shows increased soot due to a toroidal vortex around the fuel jet. While hydrogen can enhance combustion performance, it also poses challenges in terms of increased emissions and soot formation, particularly under certain injection conditions.

Xiao et al. [14] studied deflagration of hydrogen in a duct filled with fuel air mixture and perturbed by a nitrogen jet, which is transverse to the moving flame and forms a jet in crossflow configuration. They have used high speed schlieren visualization to analyze the flame's behavior and characterize the flow field. Heated hydrogen jet flame is studied in preheated conditions experimentally by Steinberg et al. [15]. Jet flames in cross flow exhibit two branches: a stable lee-stabilized branch and a dynamic lifted branch, influenced by recirculation zones and strain-rate effects. The flame position is observed to depend strongly on the fluid mechanical strain rate field. A review on hydrogen flammability [16] is useful for understanding the flame stabilization and chemical pathways. The review covers the structures and stabilization mechanisms of hydrogen diffusion flames, noting current uncertainties in extinction conditions. It also discusses deflagration instabilities and the effects of strain and curvature on flame behavior. Another study [17] involved both experimental and numerical determination of laminar burning velocities and burned Markstein lengths for various fuel blends, including mixtures of hydrocarbons with hydrogen. The issue of fuel flexibility was the focus of Lieuwen et al. [18]. They considered lean premixed as turbine operation and discussed operation envelope in terms of flashback, blowout, dynamic stability, etc. of the flame for natural has, syngas, etc. An experimental study was conducted using small multi-tube burner for hydrogen fired gas turbine applications [19]. The multi-hole nozzle utilized the jet in crossflow configuration to stabilize the flame. Nair et al. [20] have investigated reacting jet in crossflow using high speed laser diagnostics. They have varied the density ratio, momentum ratio, and rate of heat release by varying the composition of jet using helium and nitrogen blending with hydrogen. Heat release if found to affect the jet stability by suppressing the growth of shear layer vortices. Grout et al. [21] have carried out a DNS study of reacting jet in crossflow configuration. They have studied the location of heat release and discuss their role in flame stabilization. Another DNS study was reported by Kolla et al. [22] where the angle of injection is varied and its effect on flame is investigated. The impact of angle of flame stability and blowout is also probed. Sayadi and Schmidt [23] have conducted a numerical study to examine the instability modes in a reacting as well as non-reacting jet in crossflow. They have obtained frequency response of jet based on advanced numerical methods. Recently, using a numerical framework, Balaji et al. [24] have studied jet in crossflow for different gas jets. They have focused on the effect of jet density and have demonstrated that argon jet, owing to its higher density penetrates significantly higher than helium jet for the same conditions. This density effect was incorporated into momentum flux ratio and trajectory equations were derived to estimate jet penetration.

There are a few studies on reacting hydrogen jet in crossflow, but much more research is required for complete understanding of the combustion behavior in this configuration. Flashback remains one of the biggest challenges in safe operation of any hydrogen combustor. And safety remains one of the major concerns in wider acceptance of hydrogen systems [25-27]. Injecting pure fuel jet is an effective strategy to prevent flashback, but then non-uniform mixing will lead to hot spots and possible NOx production zones. As hydrogen is lighter, it will have lower momentum flux ratio as compared to hydrocarbon fuels for the same flow rate, which will result in lower penetration [24]. Both the issues can be resolved if the mixing of flame jet can be promoted. Acoustic forcing is a promising mechanism which can lead to enhanced mixing and found to be effective in non-reacting JIC configuration. It can also provide insights on the flame dynamics when subjected to thermo-acoustic oscillations [28, 29]. To the best of authors' knowledge, there is no reported work which explores the effect of acoustic forcing in hydrogen flames, especially in crossflow configuration, which is the main focus of the present study. Further, the unsteady dynamics of the forced flame is investigated using Proper Orthogonal Decomposition (POD) which is an advanced post-processing tool [30-33] used for identifying coherent structure in unsteady flow fields, and is successfully employed in the crossflow configuration [32, 33].

## 2. Experimental Facility and Visualization Technique

The experimental facility consists of air and fuel flow lines, acoustic forcing set-up, and Shadowgraphy imaging facility. All these units are described in following sub-sections.

### *2.1. Air and fuel flow lines*

Schematic of the experimental facility with air and fuel flow lines is shown in Fig. 1(a). Air flow is generated by a centrifugal blower. Power to the 10 HP motor running the blower is given through a Variaq which controls the blower rotational speed, and consequently the air flow rate. The air velocity is monitored by pitot tube measurements at the test section. The blower exit is connected to a settling chamber, which leads to a converging section. This converging section is designed with a fourth order polynomial profile to make the flow uniform and stable. The converging section is connected to the test section. The test section is 105 mm high, 50 mm wide and 300 mm long. It has quartz wall on the sides for imaging. The injector is flush-fitted at the center of the bottom wall. Hydrogen is supplied from cylinder through pipelines with control elements like pressure regulator, valves, and flame arrestor. The supply of hydrogen to the nozzle is controlled using a mass flow controller (make- Alicat, model- cori Flow).

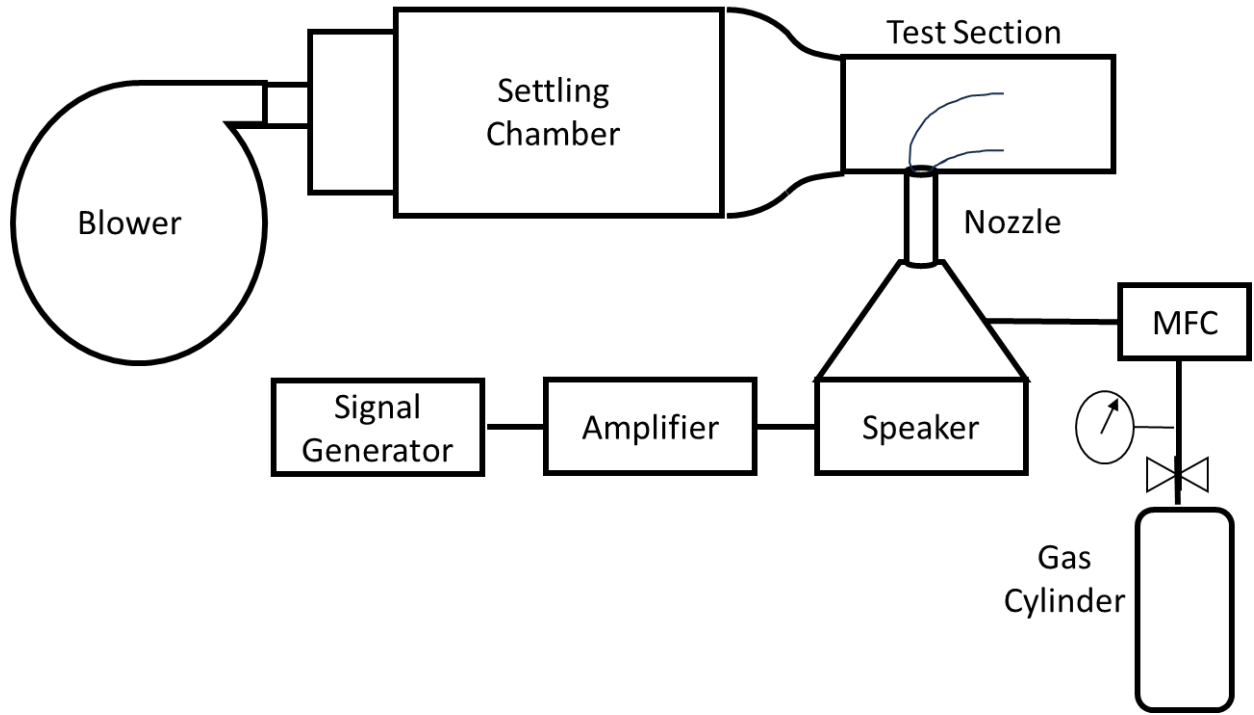

(a) Schematic of the Experimental Facility

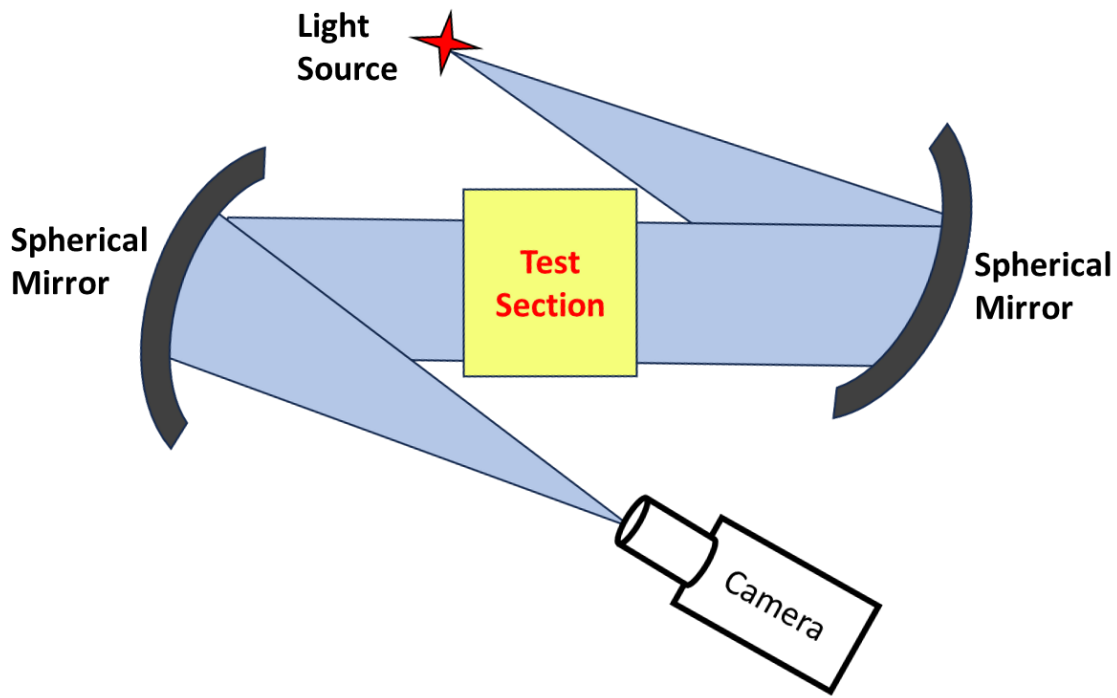

(b) Schematic of Shadowgraphy imaging technique

**Figure 1. Details of the experimental facility and Shadowgraphy imaging.**

*2.2. Acoustic forcing*

A function generator (make-Siglent, model- SIGLENT SDG-1032X) is used for acoustic forcing, which is connected to a voltage amplifier. The amplifier then controls the speaker which is encased in a conical chamber (cf. Fig. 1(a)). The chamber is sealed from all sides. It has a fuel inlet from the side wall while the top is connected to the injector. The required frequency is generated by the function generator, while the amplitude is varied with the help of the amplifier. The amplitude is characterized using a hot wire anemometer at the exit of the nozzle. Velocity measurements at the exit of the nozzle (without crossflow) are presented in Appendix A. The nozzle exit diameter is 1.2 mm.

*2.3 Shadowgraphy imaging set-up*

The schematic of the imaging set-up is shown in Fig. 1(b). Two 6-inch diameter spherical mirrors (make- Edmund optics) are placed around the test section as shown in the figure. A LED light kept in a small box with a pin-hole to act as a point source is kept at the focal length of one mirror. The light rays become parallel after getting reflected from the mirror. The parallel rays of light pass through the test section and then fall on the other spherical mirror. Reflected light from the second mirror is captured by a high-speed camera (make- Photron, model- mini AX-100) located at the focal length of the second mirror, as shown in Fig. 1(b). The density difference in the flow field in the test section contributes to bending of parallel light rays which are captured by the camera and helps to visualize the hydrogen flame. All the images captured for this study are at a frame rate of 10,000 frames per second.

*2.4. Proper Orthogonal Decomposition (POD)*

POD is an advanced data processing technique which can be used for revealing ordered structures from seemingly random flow fields. It extracts the dominant structures and their associated frequencies, and is a very powerful tool in fluid mechanics research. POD mode shapes show the dynamic structures in the flow field. While the Power Spectral Density (PSD) plots identify the frequency or range or frequencies associated with the fluid structures. Overall, it becomes very useful tool to extract useful information from a large set of experimental or computational data. POD modes are arranged in descending order of energy or significance. For details of POD analysis and its advantages and applications, the readers are referred to seminal review papers [30, 31] and examples demonstrating POD applications in crossflow configuration [32, 33].

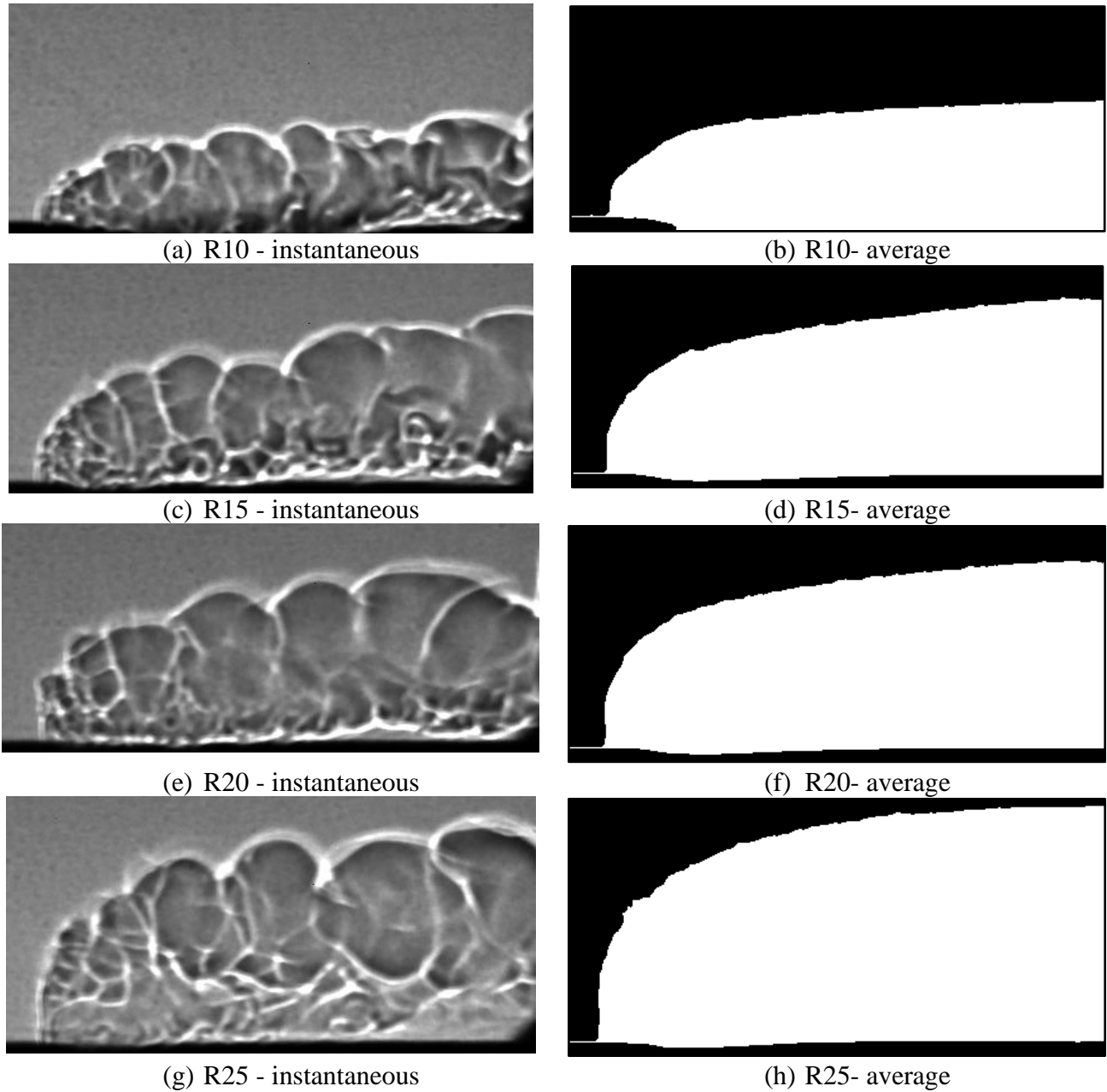

**Figure 2. Hydrogen JIC- instantaneous (left) and average (right) images for various cases**

# 3. Results and Discussion

## 3.1. Unforced cases

The hydrogen jet is injected in the crossflow as shown in Figure 2. Air velocity is maintained constant at 8.2 m/s for the entire study. Jet velocity is varied by controlling the mass flow rate of hydrogen. Average jet velocity is obtained from the mass flow rate. Velocity ratio ($R$) is the ratio of jet average velocity and duct air velocity:

$$R = \frac{U_{jet}}{U_{air}} \quad (1)$$

It is important to understand that the jet velocity only refers to the unburnt gas supply velocity. Once the fuel burns the expansion process will significantly change the velocity. However, the flame jet velocity is challenging to determine especially when it is interacting with the crossflow. Hence, the unburnt jet velocity is taken as a reference in Eq. 1. The cases with various velocity ratios are listed in Table 1. The case names are indicative of their operating conditions for easy recall while discussion.

| S No | Case | R | q |
|---|---|---|---|
| 1 | R10 | 10 | 6.8 |
| 2 | R15 | 15 | 15.4 |
| 3 | R20 | 20 | 27.3 |
| 4 | R25 | 25 | 42.7 |

Table 1. List of unforced cases with velocity ratio and momentum ratio

Another important parameter is momentum flux ratio ($q$), or simply the momentum ratio. It is expressed as the ratio of jet momentum flux to the crossflow air momentum flux:

$$q = \frac{\rho_{jet} U_{jet}^2}{\rho_{air} U_{air}^2} \quad (2)$$

Where $q$ is the momentum ratio, $\rho_{jet}$ is the jet density and $U_{jet}$ is the jet velocity respectively. Similarly, is the air density is denoted by $\rho_{air}$ and air velocity is referred to as $U_{air}$. For the present study, as both air and hydrogen have fixed density, velocity ratio is also sufficient to understand the jet dynamics.

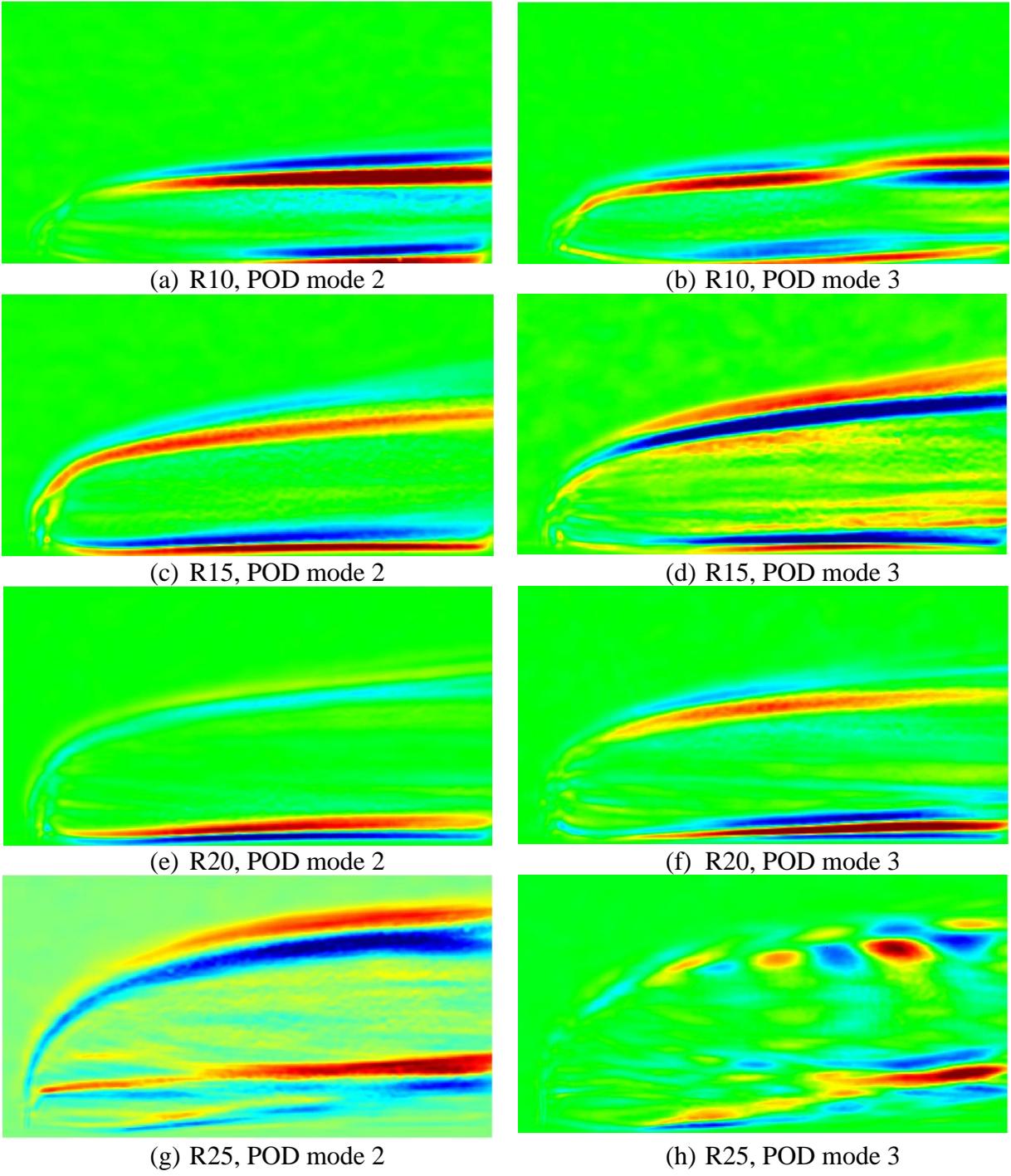

Figure 3. POD mode shapes for various unforced cases.

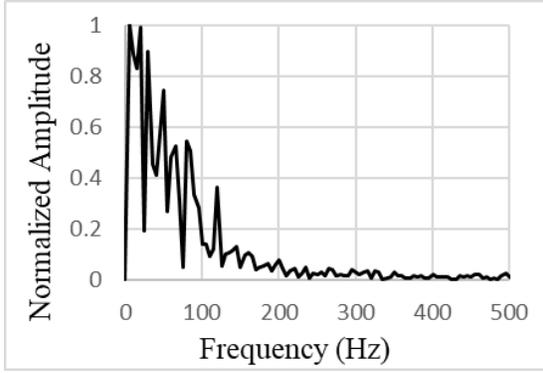
(a) R10, PSD mode 2
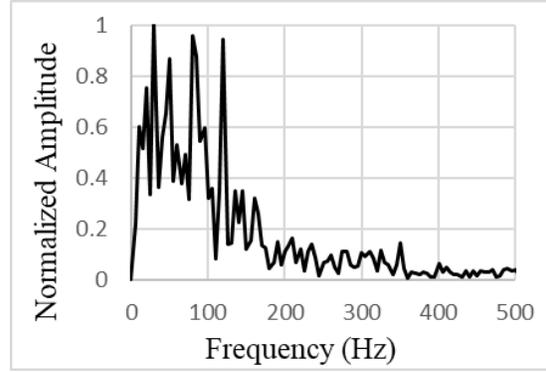
(b) R10, PSD mode 3
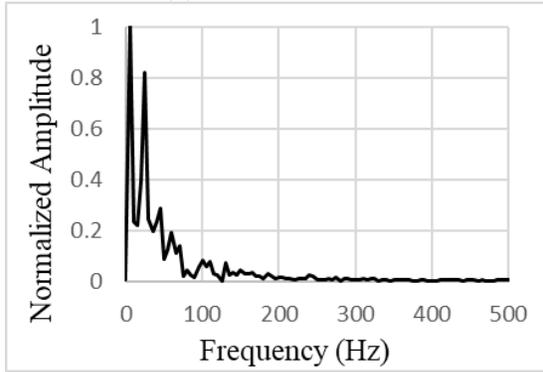
(c) R15, PSD mode 2
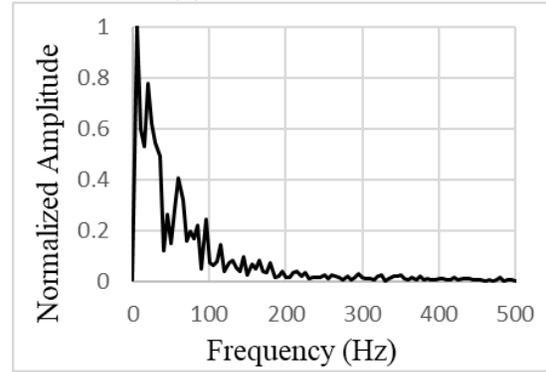
(d) R15, PSD mode 3
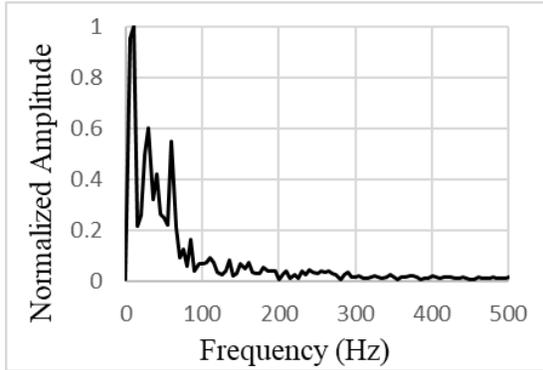
(e) R20, PSD mode 2
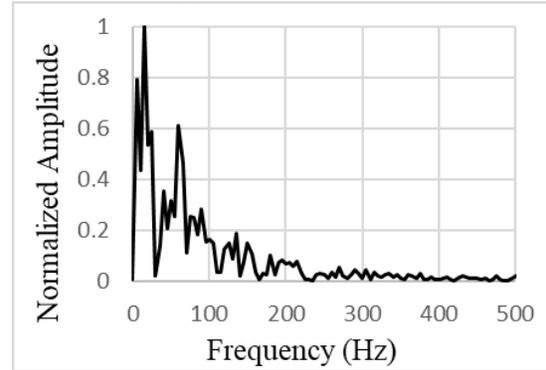
(f) R20, PSD mode 3
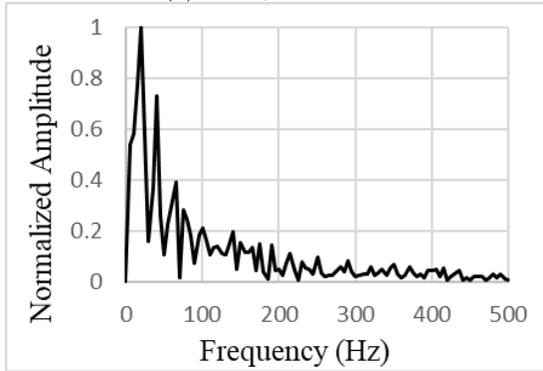
(g) R25, PSD mode 2
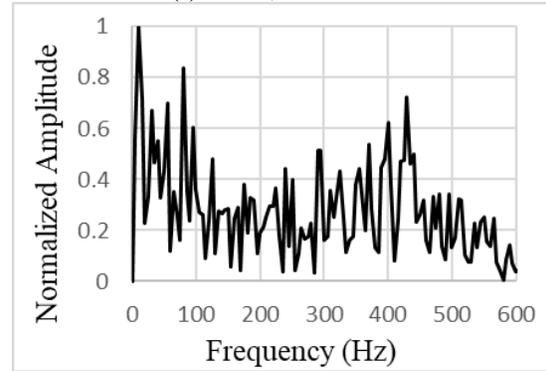
(h) R25, PSD mode 3

**Figure 4. PSD plots for various unforced cases.**

The instantaneous and averaged images of these cases are shown in Figure 2. Left column shows the instantaneous images while the corresponding averaged image is shown in the right column of Fig. 2. The flame structure and shear layer undulations are clearly visible in the instantaneous images. Several of these images are averaged and then binarized to obtain the average image which can be used to assess the jet trajectory and penetration. As evident, the jet penetration increases with increase in jet velocity, or velocity ratio ($R$). Several wrinkled structures are also visible in instantaneous flame images highlighting the rich dynamics of the flow field. The flow field is more vigorous and turbulent in the near nozzle region, where the jet first comes into contact with the crossflow, and the flame is anchored. However, as we move away from the injector, the crossflow air dominates and the initial instability dies down gradually. The shear layer formed between the flame jet and the crossflow also exhibits periodic oscillations which are convected away by the crossflow. To further gain insights, POD analysis was performed on the instantaneous images. The POD mode shapes are shown in Fig. 3. The first POD mode exhibits the average jet structure which is not shown as the focus is on dynamic behavior. Second mode is shown in the left column and third mode is shown in the right column in Fig. 3. The corresponding Power Spectral Density (PSD) plots are shown in Fig. 4. The POD modes shapes provide spatial information while PSD plots characterize the temporal behavior. Mode 2 of all the cases show bands in the shear layer which denote the shear layer oscillations. However, the corresponding PSD plots in Fig. 4 show peaks around 0 Hz, which corresponds to the average mode. For cases R10 to R20, the frequency peaks are around 5-20 Hz. Whereas for R25, peak frequency is around 20 and 40 Hz. The third mode shows near mean behavior for R15 and R20 cases where frequency peak is obtained around 5-20 Hz. PSD spectra of R10 and R25 are considerably more perturbed. R10 has several peaks between 0 and 100 Hz, while R25 case even has a smaller peak near 400 Hz. This same behavior is also seen in their POD mode shapes. POD mode 3 of R10 case shows some signs of shear layer oscillations, while R25 captures more features, especially near the jet shear layer. Another important point is that all mode 2 PSD plots are displaying low frequency response only, and there is barely any peak after 100 Hz. Mode 3 of R25 has considerably noisy signal, but it also dies down beyond 600 Hz. Hence, it can be safely assumed that the natural frequency for any flame jet is lower than 100 Hz, for the given range of parameters covered in this study.

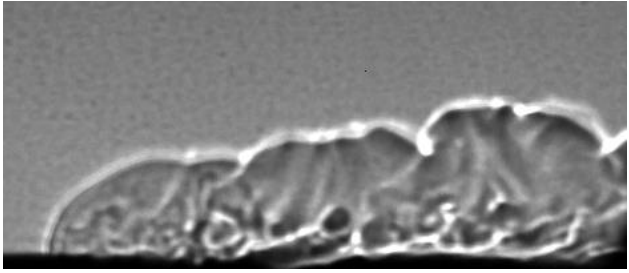
(a) R10-100

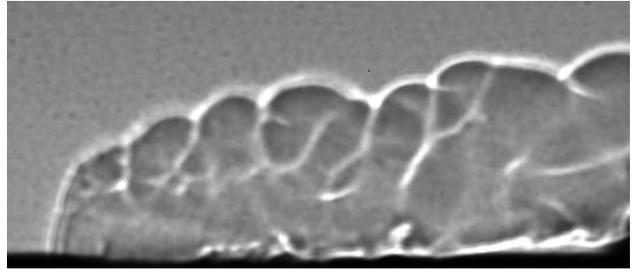
(b) R15-100

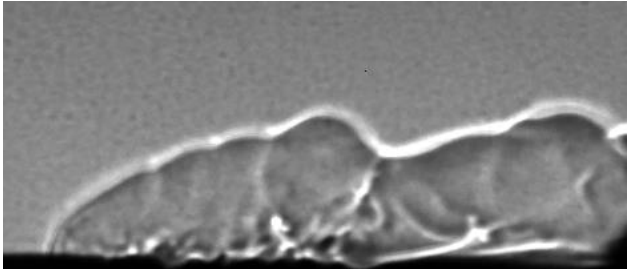
(c) R10-200

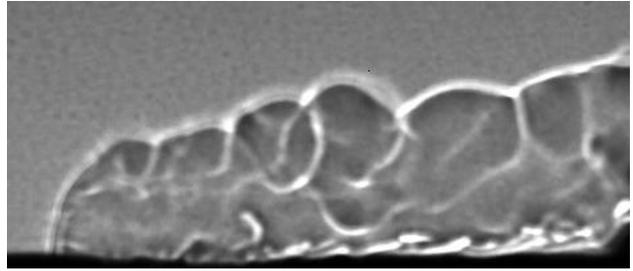
(d) R15-200

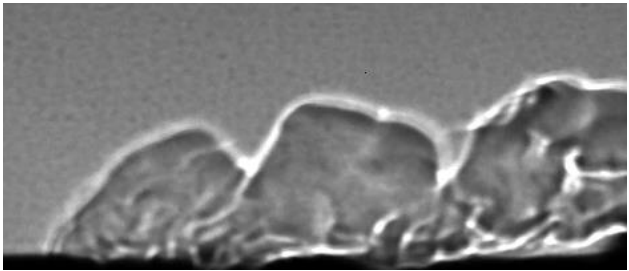
(e) R10-300

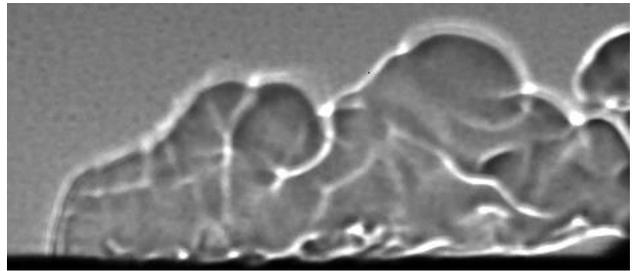
(f) R15-300

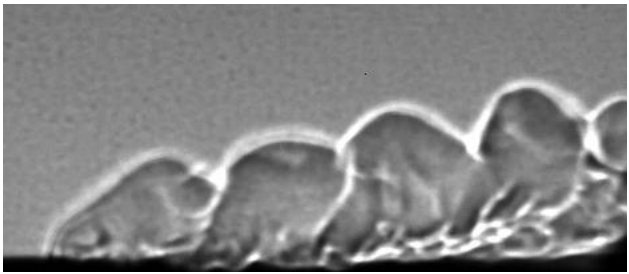
(g) R10-500

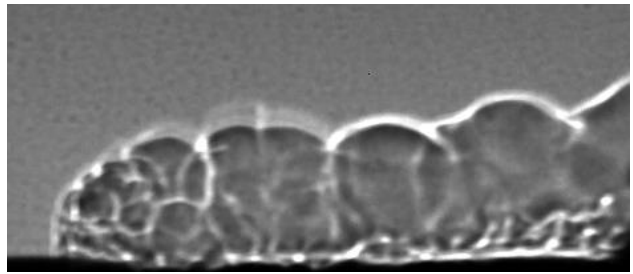
(h) R15-500

**Figure 5. Hydrogen JIC forced cases for R=10 and R=15.**

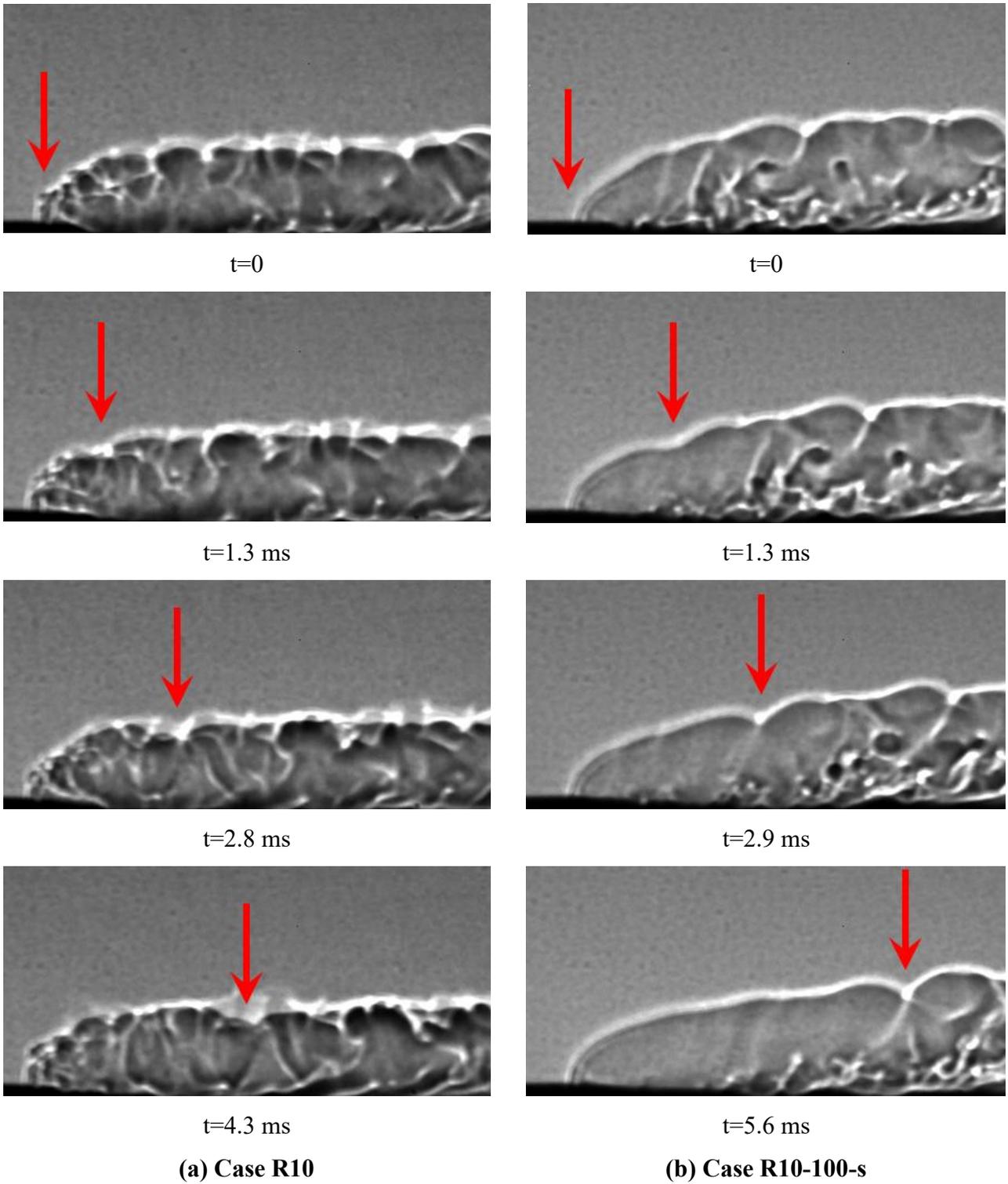

**Figure 6. Instantaneous images of (a) unforced (R10) and, (b) weakly forced (R10-100-w) case for the same velocity ratio. (time is given in milli seconds)**

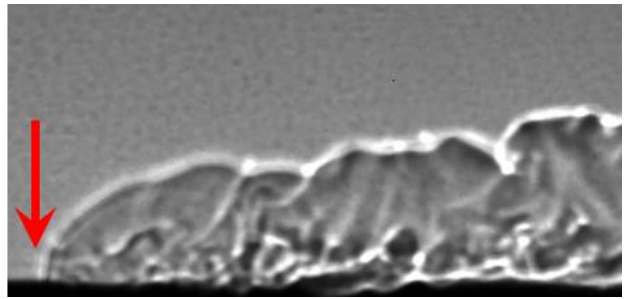
t=0

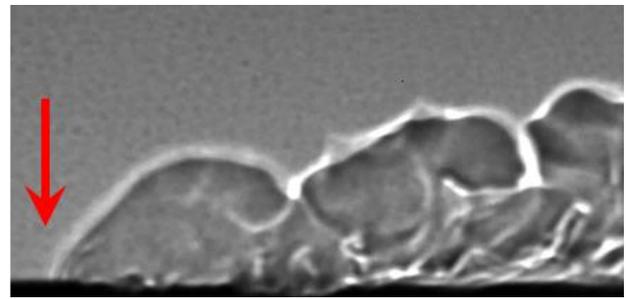
t=0

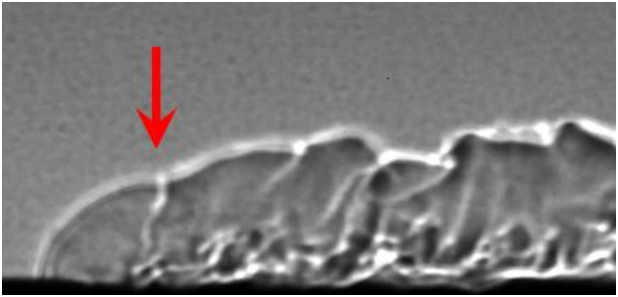
t=2 ms

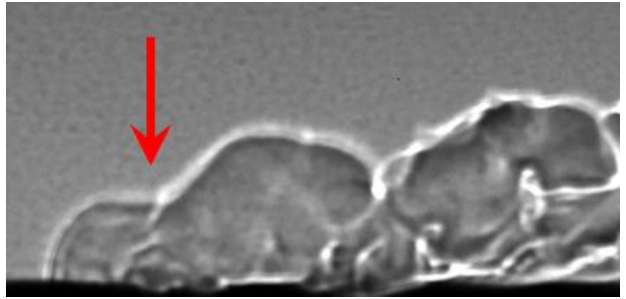
t=1.4 ms

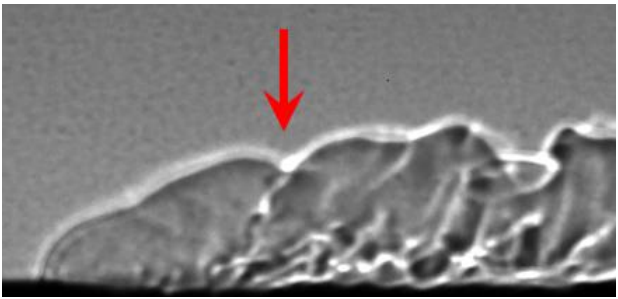
t=3.7 ms

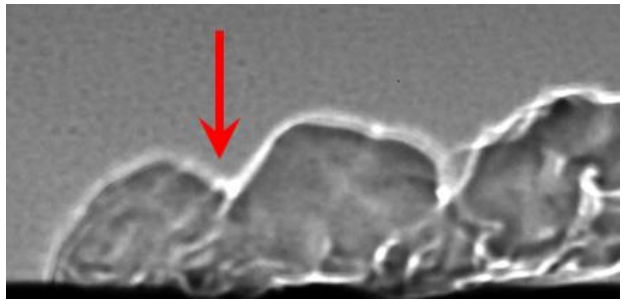
t=2.4 ms

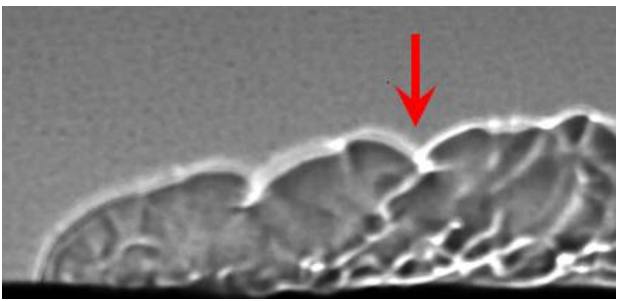
t=5.7 ms

(a) Case R10-100

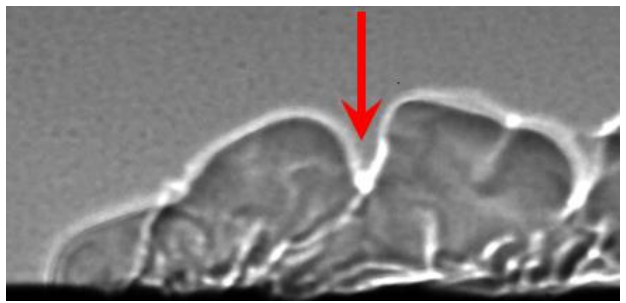
t=4.4 ms

(b) Case R10-300

**Figure 7. Instantaneous images of (a) R10-100 and, (b) R10-300 cases tracking a vortical structure (time is given in milli seconds)**

*3.2. Forced cases*

The details of forced cases are presented in Table 2. The measured velocity at the exit of nozzle is presented in Appendix A. Please note that the speaker does not respond equally to all the frequencies. For e.g., for the same function generator and amplifier settings, 100 Hz frequency shows $\alpha$ of 64%, while 500 Hz shows 11% of $\alpha$. $\alpha$ is the measure of perturbation, defined in Equation 2. Closest values of $\alpha$ practically obtained for different frequencies are used for comparison. Further, instantaneous images of forced cases are shown in Fig. 5. As these are highly unsteady cases, a singe instantaneous image is not sufficient to analyze a case. Again, POD analysis will be employed along with some sequential images to explain the processes involved. To appreciate the effect of forcing, two cases are compared in Fig. 6. Figure 6(a) shows the unforced case (R10) while Fig. 6(b) shows a weakly forced case (R10-100-s, cf. Table 2). The arrow symbols mark a typical feature in the flame which is tracked in the instantaneous images. Please note that these images are not consecutive images. They demonstrate the downstream travel of flame wrinkling structures. The unforced case seems to consist of a large number of small vortical structures which gradually traverse downstream. Interestingly, the smaller vortical structures reduce significantly in the forced flame. The jet penetration is also higher for the forced flame. Details regarding forced flame cases are given in Table 2.

| S no | Case No | Velocity Ratio (R) | Forcing Frequency (Hz) | Amplitude $\alpha$ (%) |
|---|---|---|---|---|
| 1 | R10-100-s | 10 | 100 | 22 |
| 2 | R10-100 | 10 | 100 | 64 |
| 3 | R10-200 | 10 | 200 | 29 |
| 4 | R10-300 | 10 | 300 | 31 |
| 5 | R10-500 | 10 | 500 | 11 |
| 6 | R15-100-w | 15 | 100 | 22 |
| 7 | R15-100 | 15 | 100 | 64 |
| 8 | R15-200 | 15 | 200 | 29 |
| 9 | R15-300 | 15 | 300 | 31 |
| 10 | R15-500 | 15 | 500 | 11 |

**Table 2. Details of forced JIC cases.**

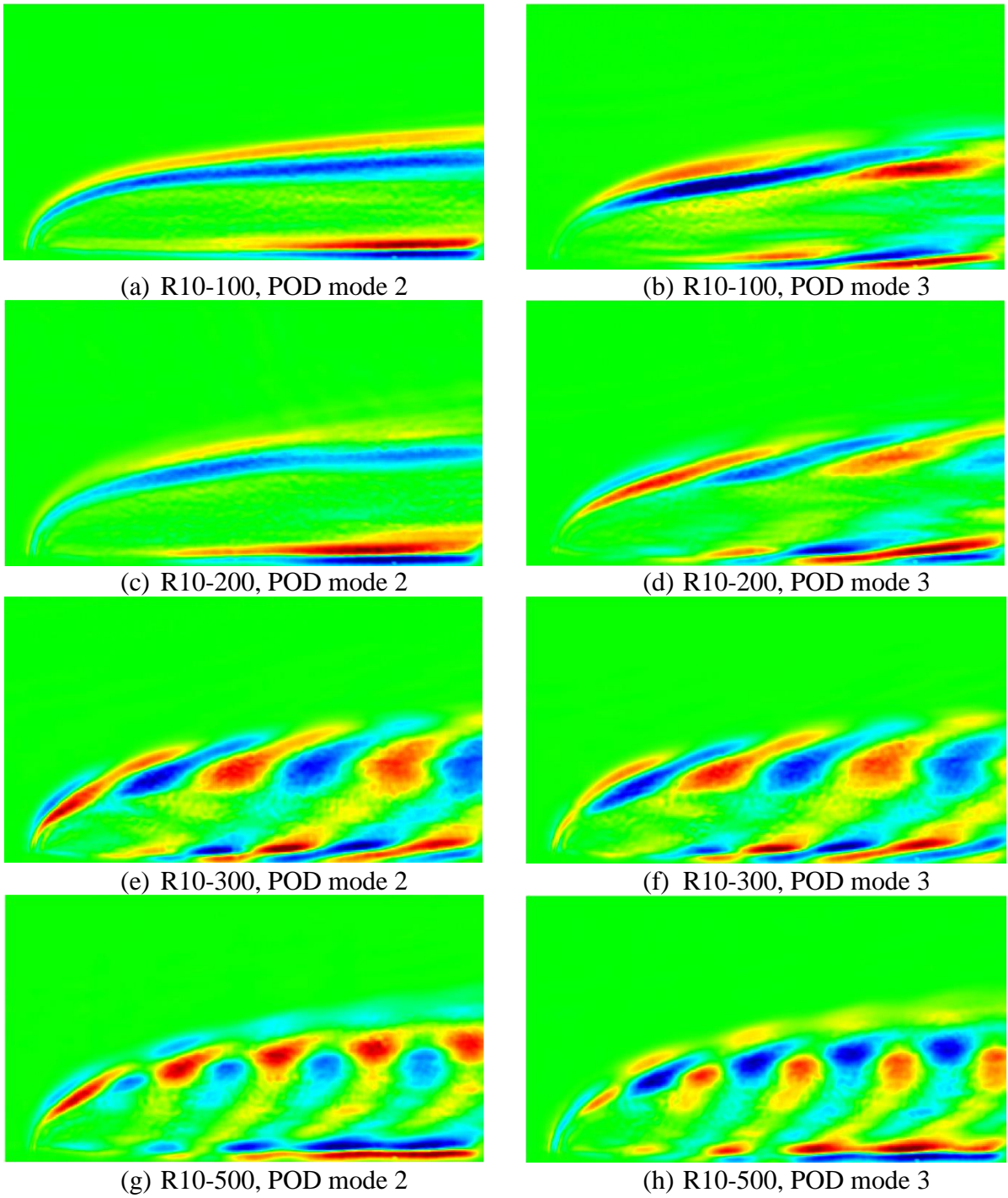

**Figure 8. POD mode shapes (mode 2 and 3) for forced cases (R=10)**

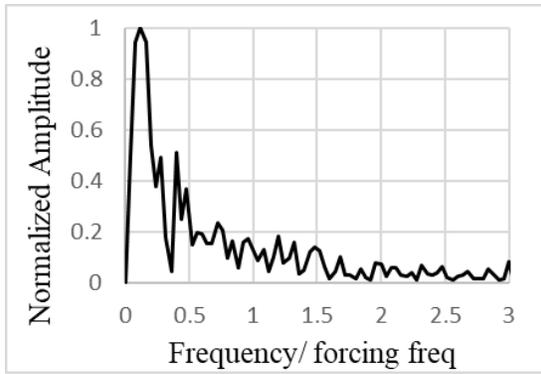
(a) R10-100, POD mode 2

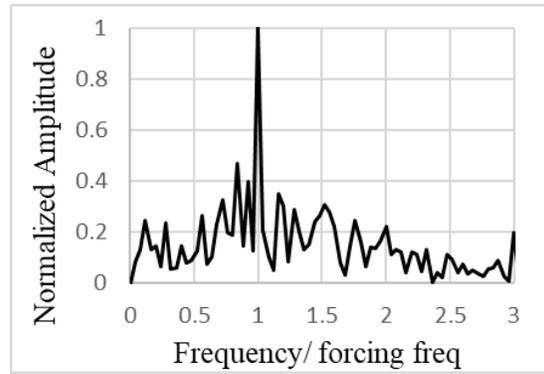
(b) R10-100, POD mode 3

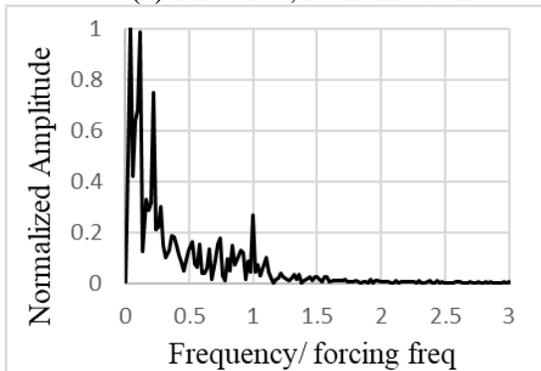
(c) R10-200, POD mode 2

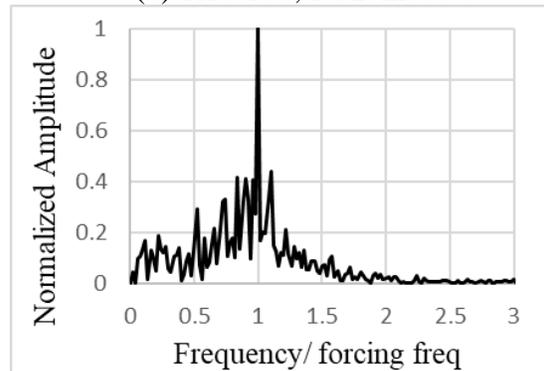
(d) R10-200, POD mode 3

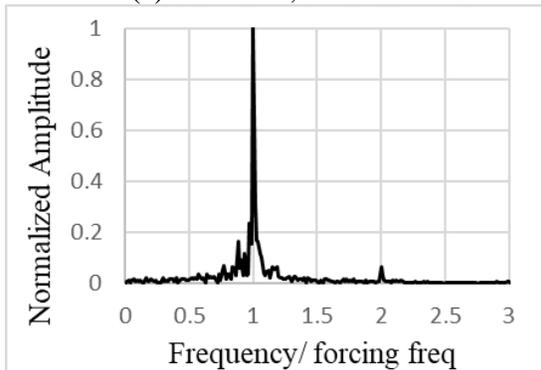
(e) R10-300, POD mode 2

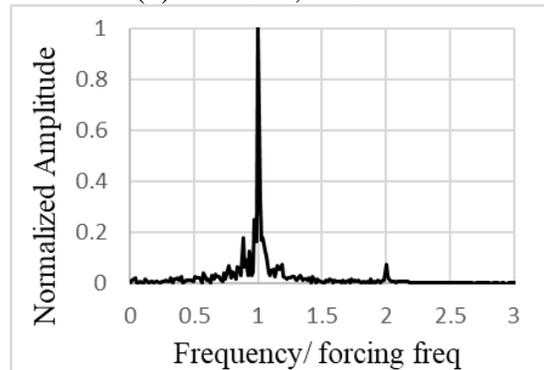
(f) R10-300, POD mode 3

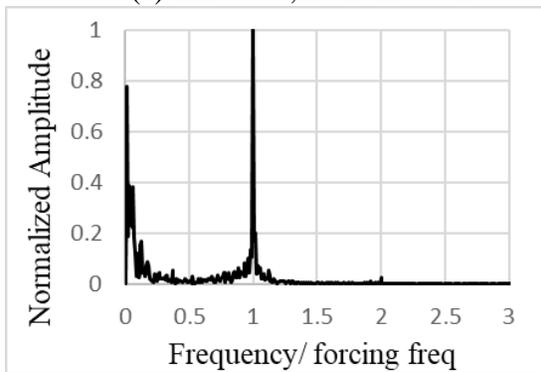
(g) R10-500, POD mode 2

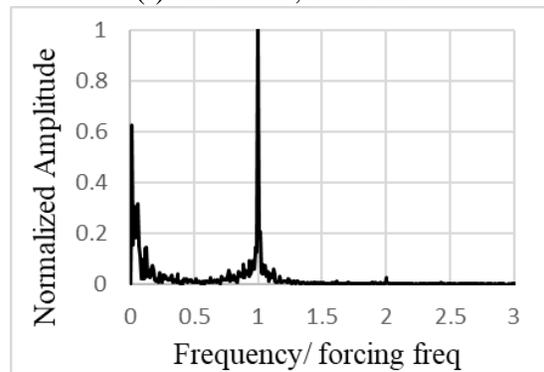
(h) R10-500, POD mode 3

**Figure 9. PSD plots for forced cases for R=10**

The forcing function can be described as:

$$U = U_0 \left(1 + \alpha \sin(\omega t)\right) \quad (2)$$

where $U$ is the instantaneous jet velocity, $U_0$ is the unperturbed velocity, $\alpha$ is the amplitude $\omega$ and is the angular frequency. Instantaneous images for cases R10-100 and R10-300 are shown in Fig. 7. Both image sequences track a vortical structure while travelling downstream. As evident, the flame shows significantly better response to the 300 Hz perturbation (R10-300) as compared to a 100 Hz (R10-100) perturbation. In spite of the fact that the forcing function of the R10-100 case has much higher $\alpha$ than the R10-300 case (cf. Table 2). To probe this observation further, POD analysis is carried out for various cases. POD mode shapes of forced cases with R=10 are presented in Fig. 8, and PSD plots for the corresponding cases are shown in Fig. 9. PSD plots for forced cases are normalized with the forcing function frequency. Mode 2 of Case R10-100 show a near mean behavior. The mode shape shows the shear layer region marked while the PSD plots show a near mean frequency. The third mode of this case, however show peak at the forcing frequency and shear layer oscillations being captured in the mode shape. Similar behavior is observed for case R10-200. Mode 2 is more significant than mode 3 in any POD result. Hence, these observations indicate suppressed response to the forcing function. Further, case R10-300 and R10-500 show much pronounced response to the forcing function. The mode shapes clearly show structures travelling downstream. These structures can be linked to vortical structures moving downstream observed in instantaneous images. The PSD plots also show a dominant peak at the forcing function frequency. It is also evident that the R10-300 case shows better response to the forcing function as it has more clearly defined structures in POD mode shapes. Case R10-300 peaks only at the forcing function frequency, while case R10-500 case also has a clear peak near 0 Hz frequency indicating the dominance of the average flow behavior in both mode 2 and 3.

Moreover, cases for R=15 are explored following the same procedure. The POD mode shapes are shown in Fig. 10 while PSD plots are presented in Fig. 11. The flame response to forcing frequency follows the same trend as observed for R=10 cases. The first two cases (R15-100 and R15-200) show dominance of mean flow in their POD mode shapes, where the observed mode shapes indicate the shear layer outline.

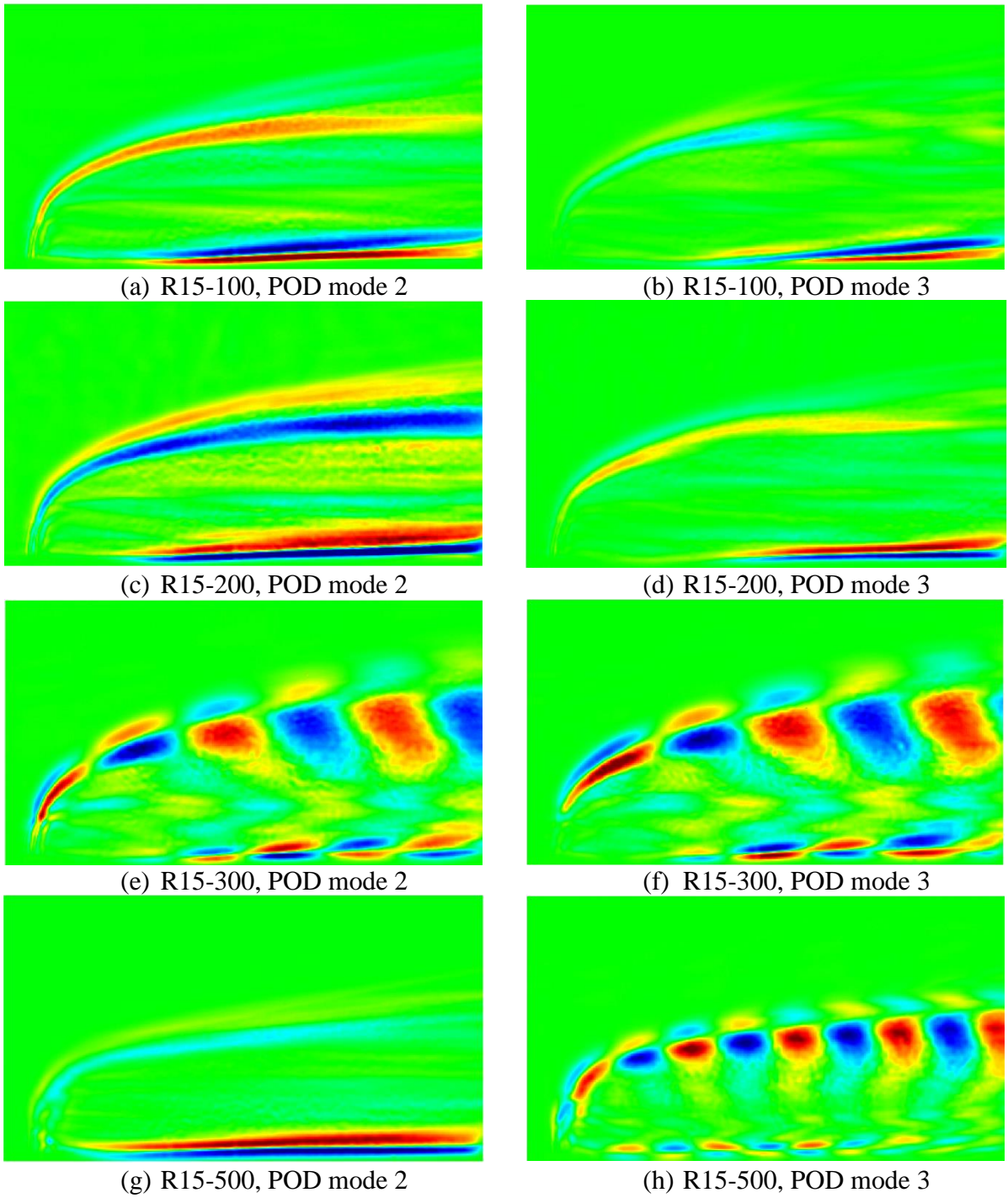

Figure 10. POD mode shapes (mode 2 and 3) for forced cases (R=15)

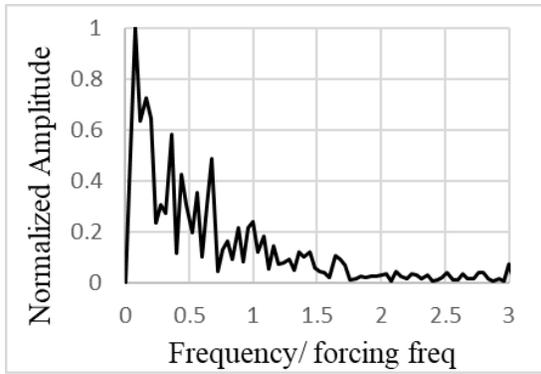
(a) R15-100, POD mode 2

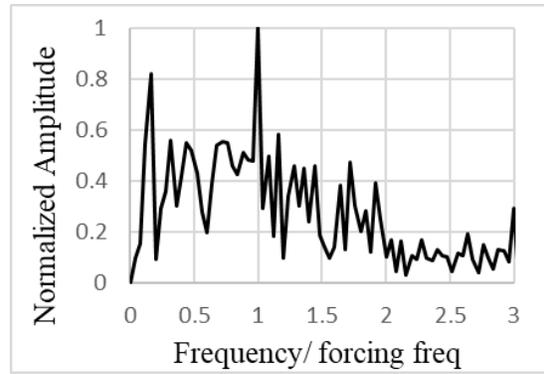
(b) R15-100, POD mode 3

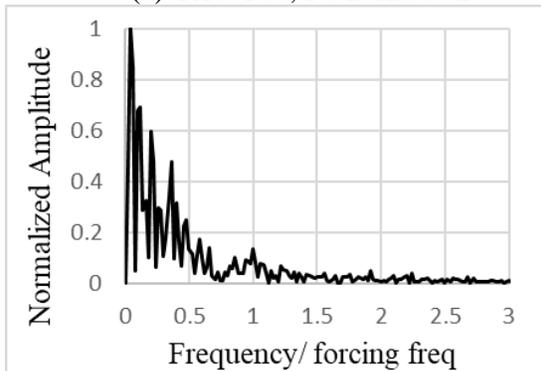
(c) R15-200, POD mode 2

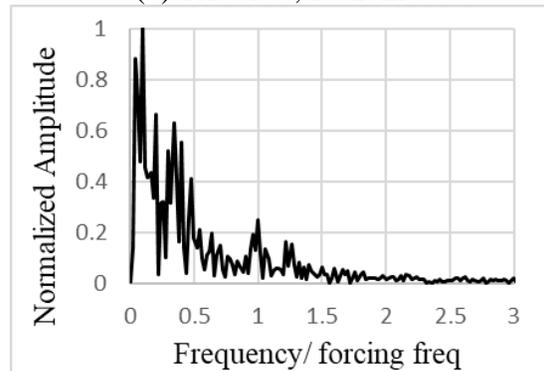
(d) R15-200, POD mode 3

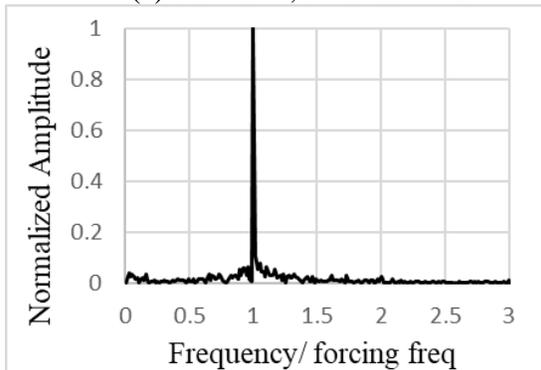
(e) R15-300, POD mode 2

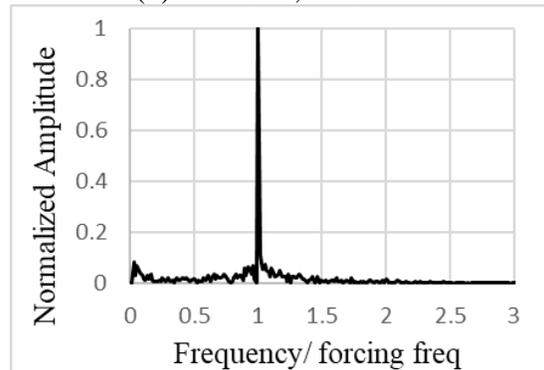
(f) R15-300, POD mode 3

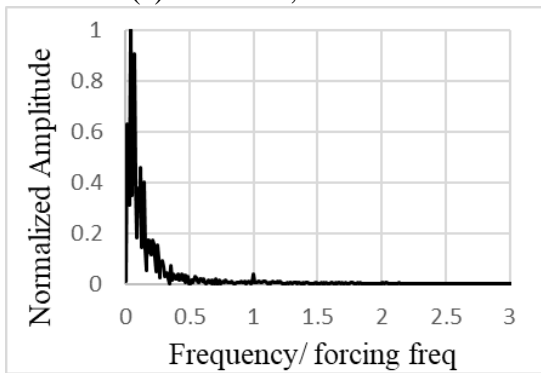
(g) R15-500, POD mode 2

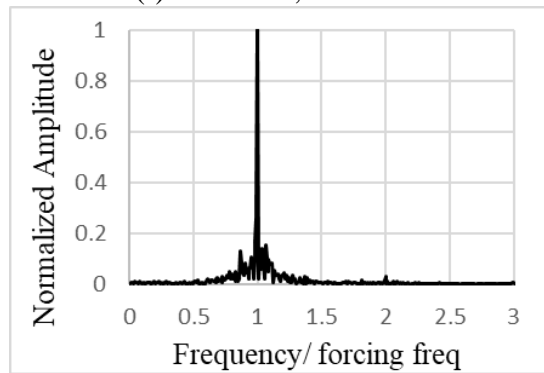
(h) R15-500, POD mode 3

**Figure 11. PSD plots for forced cases for R=15**

Their PSD plots exhibit mean behavior while being noisy. The third mode which shows forcing frequency peak for R=10 cases, is also very noisy and shows a peak for R15-100. Whereas, case R15-200 show peak near 0 Hz and appears to skewed for the mean mode behavior. Further, case R15-300 shows a distinct peak for forcing function and mode shape corresponding to the moving vortical structures which are very clearly defined. 300 Hz appears to be the most suitable frequency for both the velocity ratios investigated for this study. The preference on frequency can further be investigated using instability analysis tools. However, detailed characterization of the velocity and density, especially in the shear layer is required to do a detailed instability analysis. This lies beyond the scope of the present work and will be dealt in future studies.

Moreover, the 500 Hz case (R15-500) is surprisingly showing the mean mode behavior in PSD plots. The POD mode shapes are also faintly defined than their counterpart (R10-500) case for R=10. It can be inferred that the response to forcing function is diminishing with increase of velocity ratio from R=10 to R=15. To examine this behavior in more detail, the instantaneous images of R10-500 and R15-500 can be compared from Fig. 5. As evident, the R=15 case (Fig. 5(h)) is much perturbed and turbulent at the point of injection. This enhanced interaction of jet with crossflow possibly suppresses the forcing. Whereas, the effect of crossflow in R10-500 case (Fig. 5(g)) is less dominant and it is more receptive of the forcing function.

## 4. Conclusions

The study experimentally investigates the reacting jet of hydrogen in presence of crossflow of air. Various cases with different momentum ratios and forcing frequency combination are studied. As the hydrogen flame is usually not visible, density-based imaging is required. High speed Shadowgrpahy is used to capture unsteady jet dynamics. Since the configuration is highly dynamic in nature, single instantaneous image cannot be used for understanding the flow dynamics. A sequence of images is used to highlight any observation or differentiate behavior of two cases. Further advanced post processing tool, POD is used to analyze instantaneous images and provide spatio-temporal information of the flow field, providing a means to estimate jet natural frequency and assess response to forcing function. For unforced jet, jet penetration increases with increase in velocity ratio, which is consistent with previous studies on non-reacting jets. For forced jet, it is observed that the jet responds well to frequency in the range 300-500 Hz while the response to lower frequency was not so pronounced. This aspect needs to be dealt with a comprehensive instability analysis which will be the subject of future research.

## Acknowledgements

The authors gratefully acknowledge the support provided by Department of Science and Technology, Government of India through the SERB-SRG grant.

## References

[1] Mahesh, K. (2013). The interaction of jets with crossflow. Annual review of fluid mechanics, 45(1), 379-407.

[2] Karagozian, A. R. (2010). Transverse jets and their control. *Progress in energy and combustion science*, *36*(5), 531-553.

[3] Karagozian, A. R. (2014). The jet in crossflow. Physics of Fluids, 26(10).

[4] Murugavel, A. B., Massey, J. C., Tanaka, Y., & Swaminathan, N. (2024). The effect of methane addition on reacting hydrogen jets in crossflow. *International Journal of Hydrogen Energy*, *80*, 57-67.

[5] Sharma, V., Tang, Y., & Raman, V. (2024). Stabilization of Hydrogen-enriched Jet Flames in a Crossflow. In *AIAA SCITECH 2024 Forum* (p. 0393).

[6] Vaysse, N., Durox, D., Vicquelin, R., Candel, S., & Renaud, A. (2023, June). Stabilization and Dynamics of Pure Hydrogen Swirling Flames Using Cross-Flow Injection. In *Turbo Expo: Power for Land, Sea, and Air* (Vol. 86953, p. V03AT04A043). American Society of Mechanical Engineers.

[7] Rasheed, I., & Mishra, D. P. (2023). Numerical study of transverse hydrogen injection in high-speed reacting crossflow. *Physics of Fluids*, *35*(6).

[8] Boxx, I., Pareja Restrepo, J., & Saini, P. (2022, May). An Experimental Study of Natural Gas Jet-flames in Crossflow with High Hydrogen Admixture at Elevated Pressure Conditions. In *20th International Symposium on Applications of Laser and Imaging Techniques to Fluid Mechanics*.

[9] Saini, P., Chterev, I., Pareja, J., Aigner, M., & Boxx, I. (2021). Effects of Hydrogen-Enrichment on Flame-Holding of Natural Gas Jet Flames in Crossflow at Elevated Temperature and Pressure. *Flow, Turbulence and Combustion*, *107*(1), 219-243.

[10] Saini, P., Chterev, I., Pareja, J., Aigner, M., & Boxx, I. (2020). Effect of pressure on hydrogen enriched natural gas jet flames in crossflow. *Flow, Turbulence and Combustion*, *105*, 787-806.

[11] Xu, C., Ameen, M., Pal, P., & Som, S. (2022). Direct Numerical Simulation of a Reacting Hydrogen Jet in Turbulent Vitiated Crossflow Using Spectral Element Method. In *AIAA SCITECH 2022 Forum* (p. 0823).

[12] Zhao, M., Ye, T., & Li, Q. (2021). Large-Eddy simulations of transverse jet mixing and flame stability in supersonic crossflow. *AIAA Journal*, *59*(6), 2126-2142.

[13] Olivani, A., & Cozzi, F. (2006). Effects of Hydrogen Addition on a Confined Lean Non-Premixed Natural Gas Swirled Flame. In *HYSYDAYS 1st World Congress of Young Scientists on Hydrogen Energy Systems*. Begell House.

[14] Xiao, Q., Cheng, J., Zhang, B., Zhou, J., & Chen, W. (2021). Schlieren visualization of the interaction of jet in crossflow and deflagrated flame in hydrogen-air mixture. *Fuel*, *292*, 120380.

[15] Steinberg, A. M., Sadanandan, R., Dem, C., Kutne, P., & Meier, W. (2013). Structure and stabilization of hydrogen jet flames in cross-flows. *Proceedings of the Combustion Institute*, *34*(1), 1499-1507.

[16] Sánchez, A. L., & Williams, F. A. (2014). Recent advances in understanding of flammability characteristics of hydrogen. *Progress in Energy and Combustion Science*, *41*, 1-55.

[17] Bouvet, N., Halter, F., Chauveau, C., & Yoon, Y. (2013). On the effective Lewis number formulations for lean hydrogen/hydrocarbon/air mixtures. *International journal of hydrogen energy*, *38*(14), 5949-5960.

[18] Lieuwen, T., McDonell, V., Petersen, E., & Santavicca, D. (2008). Fuel flexibility influences on premixed combustor blowout, flashback, autoignition, and stability.

[19] York, W. D., Ziminsky, W. S., & Yilmaz, E. (2013). Development and testing of a low NOx hydrogen combustion system for heavy-duty gas turbines. *Journal of Engineering for Gas Turbines and Power*, *135*(2), 022001.

[20] Nair, V., Wilde, B., Emerson, B., & Lieuwen, T. (2019). Shear layer dynamics in a reacting jet in crossflow. *Proceedings of the Combustion Institute*, *37*(4), 5173-5180.

[21] Grout, R. W., Gruber, A., Yoo, C. S., & Chen, J. H. (2011). Direct numerical simulation of flame stabilization downstream of a transverse fuel jet in cross-flow. *Proceedings of the Combustion Institute*, *33*(1), 1629-1637.

[22] Kolla, H., Grout, R. W., Gruber, A., & Chen, J. H. (2012). Mechanisms of flame stabilization and blowout in a reacting turbulent hydrogen jet in cross-flow. *Combustion and Flame*, *159*(8), 2755-2766.

[23] Sayadi, T., & Schmid, P. J. (2021). Frequency response analysis of a (non-) reactive jet in crossflow. *Journal of Fluid Mechanics*, *922*, A15.

[24] Balaji, S., Kumar, D., Parasuram, I. V. L. N., & Sinha, A. (2021, December). Transverse Gas Jet Injection—Effect of Density Ratio. In Conference on Fluid Mechanics and Fluid Power (pp. 245-249). Singapore: Springer Nature Singapore.

[25] Sinha, A., & Wen, J. X. (2019). A simple model for calculating peak pressure in vented explosions of hydrogen and hydrocarbons. International journal of hydrogen energy, 44(40), 22719-22732.

[26] Sinha, A., Rao, V. C. M., & Wen, J. X. (2019). Modular phenomenological model for vented explosions and its validation with experimental and computational results. Journal of Loss Prevention in the Process Industries, 61, 8-23.

[27] Sinha, A., & Wen, J. X. (2018, August). Phenomenological modelling of external cloud formation in vented explosions. In 12th international symposium on hazards, prevention, and mitigation of industrial explosions (ISHPMIE).

[28] Lieuwen, T. C. (2012). Unsteady Combustor Physics. (No Title).

[29] Lieuwen, T. C., & Yang, V. (Eds.). (2005). Combustion instabilities in gas turbine engines: operational experience, fundamental mechanisms, and modeling. American Institute of Aeronautics and Astronautics.

[30] Taira, K., Brunton, S. L., Dawson, S. T., Rowley, C. W., Colonius, T., McKeon, B. J., ... & Ukeiley, L. S. (2017). Modal analysis of fluid flows: An overview. Aiaa Journal, 55(12), 4013-4041.

[31] Taira, K., Hemati, M. S., Brunton, S. L., Sun, Y., Duraisamy, K., Bagheri, S., ... & Yeh, C. A. (2020). Modal analysis of fluid flows: Applications and outlook. AIAA journal, 58(3), 998-1022.

[32] Sinha, A., & Ravikrishna, R. V. (2019). Experimental studies on structure of airblast spray in crossflow. *Sādhanā*, *44*(5), 113.

[33] Sinha, A. (2023). Effect of injector geometry in breakup of liquid jet in crossflow–insights from POD. *International Journal of Multiphase Flow*, *167*, 104497.

# Appendix A

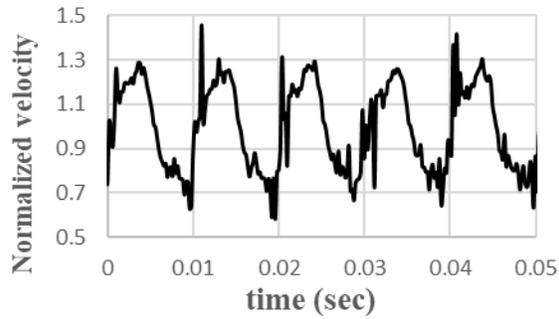
(a) 100 Hz-s (lower $\alpha$)

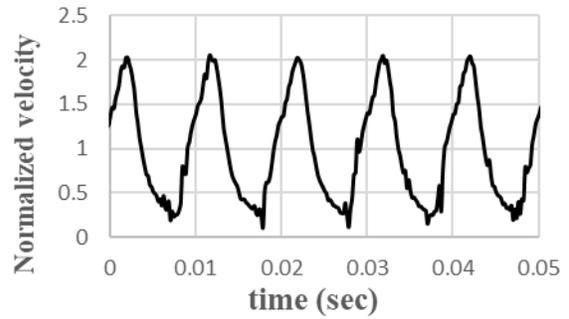
(b) 100 Hz

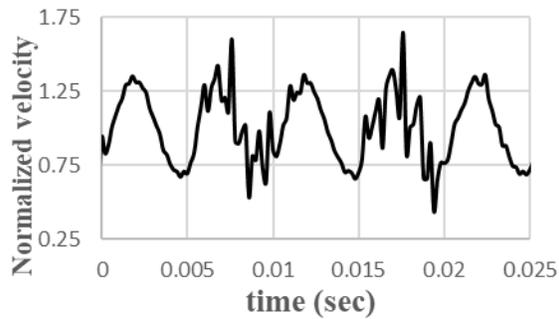
(c) 200 Hz

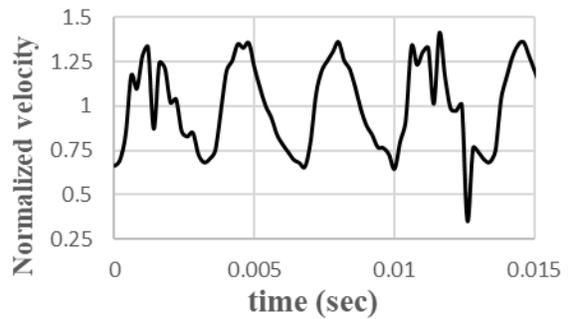
(d) 300 Hz

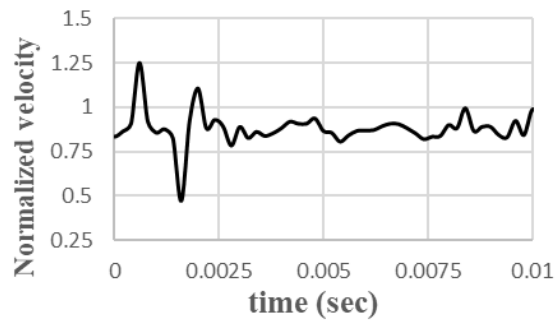
(e) 500 Hz

**Figure A1. Normalized velocity plots measured from hot wire anemometer for various forcing frequencies at the exit of the nozzle (without crossflow)**